\documentclass[12pt]{article}

\usepackage{cite}

\usepackage{amssymb,latexsym}

\usepackage{color}

\let\a=\alpha \let\b=\beta \let\g=\gamma \let\d=\delta \let\e=\epsilon
  \let\th=\theta  \let\k=\kappa
\let\l=\lambda \let\m=\mu \let\n=\nu \let\x=\xi \let\p=\pi 
\let\s=\sigma     
 
        \let\Th=\Theta 
\let\X=\Xi  \let\S=\Sigma  \let\Y=\Psi
 
\let\la=\label  
  
\def\nn{\nonumber} \def\bd{\begin{document}} \def\ed{\end{document}}
\def\ds{\documentstyle} \let\fr=\frac \let\bl=\bigl \let\br=\bigr
\let\Br=\Bigr \let\Bl=\Bigl
\let\bm=\bibitem
\let\na=\nabla
\def\tU{{\widetilde U}}
\let\pa=\partial \let\ov=\overline
\def\ie{{\it i.e.\ }}
\newcommand{\be}{\begin{equation}}
\newcommand{\ee}{\end{equation}}
\def\ba{\begin{array}}
\def\ea{\end{array}}
\def\ft#1#2{{\textstyle{{\scriptstyle #1}\over {\scriptstyle #2}}}}
\def\fft#1#2{{#1 \over #2}}
\def\F#1#2{{ F_{#1}^{(#2)} }}
\def\cF#1#2{{ {\cal F}_{#1}^{(#2)} }}

\def\R{{\bf R}}
\def\sst#1{{\scriptscriptstyle #1}}
\def\oneone{\rlap 1\mkern4mu{\rm l}}
\def\e7{E_{7(+7)}}
\def\td{\tilde}
\def\wtd{\widetilde}
\def\im{{\rm i}}
\def\bog{Bogomol'nyi\ }
\newcommand{\ho}[1]{$\, ^{#1}$}
\newcommand{\hoch}[1]{$\, ^{#1}$}
\newcommand{\bea}{\begin{eqnarray}}
\newcommand{\eea}{\end{eqnarray}}
\newcommand{\ra}{\rightarrow}
\newcommand{\lra}{\longrightarrow}
\newcommand{\Lra}{\Leftrightarrow}
\newcommand{\ap}{\alpha^\prime}
\newcommand{\bp}{\tilde \beta^\prime}
\newcommand{\cB}{{\cal B}}
\newcommand{\cO}{{\cal O}}
\newcommand{\vecx}{\vec{x}}
\newcommand{\vecy}{\vec{y}}
\newcommand{\vecp}{\vec{p}}
\newcommand{\vecq}{\vec{q}}
\newcommand{\tr}{{\rm tr} }
\newcommand{\Tr}{{\rm Tr} }
\newcommand{\NP}{Nucl. Phys. }

\newcommand{\cL}{{\cal L}}
\newcommand{\cA}{{\cal A}}
\newcommand{\cT}{{\cal T}}
\newcommand{\cD}{{\cal D}}

\def\sst#1{{\scriptscriptstyle #1}}
\def\0{{\sst{(0)}}}
\def\1{{\sst{(1)}}}
\def\2{{\sst{(2)}}}
\def\3{{\sst{(3)}}}
\def\4{{\sst{(4)}}}
\def\5{{\sst{(5)}}}
\def\6{{\sst{(6)}}}
\def\7{{\sst{(7)}}}
\def\8{{\sst{(8)}}}
\def\ve{\varepsilon}
\def\vf{\varphi}
\def\F{\Phi}
\def\wg{\wedge}

\def\thb{\bar{\theta}}
\def\Thb{\bar{\Theta}}
\def\barp{\bar{p}}
\def\barq{\bar{q}}
\def\barc{\bar{c}}
\def\bard{\bar{d}}
\def\e{\epsilon}

\def \bi{\bibitem}
\def \la {\label}

\def \l {\lambda}
\def\foot{\footnote}
\def \tl  {{\tilde \l}}
\def \sql {{\sqrt \l}}
\def \adss {$AdS_5 \times S^5$\ }
\newcommand{\rf}[1]{(\ref{#1})}
\def \ov {\over}

\def\th{\theta}
\def\Th{\Theta}
\def\vth{\vartheta}
\def\btheta{{\bar\theta}}
\def\ttheta{{{\tilde\theta}}}
\def\bttheta{{{\bar\ttheta}}}
\def\vth{\vartheta}

\def\ra{\rightarrow}
\def\N{{\cal N}}
\def\F{{\cal F}}
\def\uM{\underline{M}}
\def\uA{\underline{A}}
\def\uN{\underline{N}}
\def\uP{\underline{P}}
\def\ua{\underline{a}}
\def\ub{\underline{b}}
\def\uc{\underline{c}}
\def\ud{\underline{d}}
\def\ue{\underline{e}}
\def\uf{\underline{f}}
\def\ui{\underline{i}}
\def\uj{\underline{j}}
\def\uk{\underline{k}}
\def\ual{\underline{\alpha}}
\def\ube{\underline{\beta}}
\def\um{\underline{m}}
\def\un{\underline{n}}
\def\umu{\underline{\mu}}
\def\unu{\underline{\nu}}
\def\ula{\underline{\l}}
\def\uka{\underline{\k}}
\def\usi{\underline{\s}}
\def\urh{\underline{\r}}
\def\cc{\circ}
\def\eqv{\equiv}

\def\ni{\noindent}

\def\Ep{E^{{}^{(+)}}}
\def\Em{E^{{}^{(-)}}}

\def\Mp{M^{{}^{(+)}}}
\def\Mm{M^{{}^{(-)}}}

\def \ha{{1\ov 2}}

\def\r{\rho}

\def\Y{{\rm Y}}
\def\X{{\rm X}}
\def\tY{\tilde{\rm Y}}
\def\tX{\tilde{\rm X}}
\def\dY{\dot{\rm Y}}
\def\dX{\dot{\rm X}}

\def \J {\mathcal{J}}
\def \del {\partial}

\def\dF{\dot{F}}
\def\dG{\dot{G}}
\def\df{\dot{f}}
\def \E {{\cal E}}
\def \S {{\cal S}}
\def \J {{\cal J}}

\def\ms{\mathcal{S}}
\def\mj{\mathcal{J}}
\def\soj{\fr{\ms}{\mj}}
\def \R {{\bf R}}
\def \om {\omega}
\def \bE {\bar E}
\def \x {{\cal X}}

\def \bi{\bibitem}
\def \la {\label}

\def \l {\lambda}
\def\foot{\footnote}
\def \tl  {{\tilde \l}}
\def \sql {{\sqrt \l}}
\def \adss {$AdS_5 \times S^5$\ }
\def \ov {\over}

\def \varpi {{\rm w}}

\def\thb{\bar{\theta}}
\def\Thb{\bar{\Theta}}
\def\mb{\bar{\m}}
\def\ab{\bar{\a}}
\def\zb{\bar{z}}
\def\psib{\bar{\psi}}
\def\barp{\bar{p}}
\def\barq{\bar{q}}
\def\barc{\bar{c}}
\def\bard{\bar{d}}
\def\e{\epsilon}
\def\wb{\bar{w}}
\def\lb{\bar{\l}}
\def\Jb{\bar{J}}
\def\Nb{\bar{N}}
\def\Zb{\bar{Z}}
\def\pab{\bar{\pa}}

\def\At{\tilde{A}}
\def\Bt{\tilde{B}}
\def\Ct{\tilde{C}}
\def\Dt{\tilde{D}}
\def\Et{\tilde{E}}
\def\Ft{\tilde{F}}
\def\Gt{\tilde{G}}
\def\Ht{\tilde{H}}
\def\Mt{\tilde{M}}
\def\at{\tilde{a}}
\def\bt{\tilde{b}}
\def\ct{\tilde{c}}
\def\dt{\tilde{d}}
\def\et{\tilde{e}}
\def\ft{\tilde{f}}
\def\gt{\tilde{g}}
\def\mt{\tilde{\mu}}
\def\nt{\tilde{\nu}}

\def\bA{{\bf A}}

\def\ola{\overleftarrow}
\def\ora{\overrightarrow}
\def\alt{\tilde{\a}}

\def\eh{\hat{e}}
\def\eph{\hat{\e}}
\def\ph{\hat{p}}
\def\alh{\hat{\a}}
\def\beh{\hat{\b}}
\def\gah{\hat{\g}}
\def\muh{\hat{\m}}
\def\nuh{\hat{\n}}
\def\thh{\hat{\th}}
\def\dh{\hat{d}}
\def\ih{\hat{i}}
\def\jh{\hat{j}}
\def\kh{\hat{k}}
\def\deh{\hat{\d}}
\def\wh{\hat{w}}
\def\lah{\hat{\l}}
\def\Ah{\hat{A}}
\def\Ch{\hat{C}}
\def\Omh{\hat{\Omega}}

\def\ps{\rlap{\, /}\;\,p }
\def\ks{\rlap{\, /}\;\,k }

\def\gym{g_{YM}}

\def\adot{\dot{a}}
\def\bdot{\dot{b}}
\def\bpa{\bar{\pa}}

\def\ssk{\medskip}

\renewcommand{\theequation}{\thesection.\arabic{equation}}

\renewcommand\baselinestretch{1.2}

\begin{document}

\thispagestyle{empty}

\begin{center}
{\large \bf
Near-Extremal Black Branes with $n^3$ Entropy Growth}

\vspace{1.5cm}

{\bf E. Hatefi$^{\flat}$, A.J.
Nurmagambetov$^{\natural}$ and I.Y. Park$^{\sharp}$}
\vspace{1.5cm} \\ {\it
$^{\flat}$ Abdus Salam International Centre for Theoretical Physics,}
\\ {\it Strada Costiera 11, Trieste, Italy}
\\ {\tt ehatefi@ictp.it}
\vspace{0.3cm}\\  {\it
$^{\natural}$ A.I. Akhiezer Institute for Theoretical Physics of NSC KIPT},
\\ {\it Kharkov, UA 61108, Ukraine }
\\{\tt ajn@kipt.kharkov.ua}
\vspace{0.3cm}\\
$^{\sharp}$ {\it Department of Natural and physical Sciences, Philander Smith College,}
\\
{\it Little Rock, AR 72223, USA}\\
{\tt inyongpark05@gmail.com}

\end{center}

\medskip
\medskip
\bigskip

\bigskip

\begin{abstract}
In this paper, which is a companion to ArXiv:1204.2711[hep-th], we formulate a criterion for constructing 11D/10D maximal supergravity black-brane solutions with the near-extremal $S\sim n^3T^5$ entropy-temperature relation. We present explicit examples of such solutions with special attention paid to thermodynamics of black M-waves (KK-waves), intersecting/interpolating black branes with KK-waves, M2/M5 branes and their descents. We find the conditions on charges and numbers of branes in intersections, under which the dielectric effect of brane polarization becomes manifest on the supergravity side. We also briefly discuss the shear viscosity per entropy bound of M5 and D4 black branes within the AdS/CFT hydrodynamical limit.

\end{abstract}



\newpage

\section{Introduction}

One of the most intriguing puzzles of M-theory relates to counting its degrees of freedom. Sharply non-perturbative in nature, M-theory involves various non-perturbative, i.e., solitonic degrees of freedom. They range from higher spin fields of a hidden symmetry of M-theory \cite{West:2000ga,West:2001as}, and ``preonic'' constituents \cite{Bandos:2001pu} of M-algebra \cite{Sezgin:1996cj}, to branes \cite{Englert:2003py}.
It has been hoped that 6D $(2,0)$ theory, a mysterious theory associated with the effective worldvolume description of M5 branes \cite{Howe:1996yn,Pasti:1997gx,Bandos:1997ui,Aganagic:1997zq,Howe:1997fb,Bandos:1997gm}, may provide a clue to the existence of new hidden non-perturbative constituents of M-theory \cite{Kim:2011mv,Lee:2000hp,Bolognesi:2011rq,Bolognesi:2011nh}.

As widely known, the entropy of $n$ coincident M5 branes scales as $n^3$ \cite{Klebanov:1996un}.
The difference between the near-extremal entropy growth of coincident black $n$ D3-branes and that of coincident black $n$ M5 branes was emphasized. For the near-extremal D3-branes, $S\sim n^2 T^3$ in agreement with $U(n)$ symmetry of $n$ coincident D3-branes \cite{Gubser:1996de}. For $n$ near-extremal M5 branes, the corresponding relation takes $S\sim n^3 T^5$; the $n^3$ entropy growth cannot be associated with any classical Lie group.\footnote{ A system of $n$ coincident D4 branes also exhibits $n^3$ near-extremal entropy behavior. See \cite{Ryang:1999bh} for early notice of this phenomenon.}
It was taken as an indication of novel degrees of freedom of 6D (2,0) theory.

Part of the hidden constituents of 6D $(2,0)$ theory may be related to the self-dual strings \cite{Howe:1997ue}, living on the world-volume of the M5 branes (see \cite{Berman:2007bv} for a review). They contribute to the effective potential of 6D chiral supersymmetric theory via string junctions \cite{Berenstein:1998rr,Bolognesi:2011rq} and instanton transitions \cite{Kim:2011mv, Lee:2000hp,Bolognesi:2011nh}. The $n^3$ contribution of the internal degrees of freedom was also viewed from anomaly inflow in 11D supergravity in the presence of dynamical M5 branes \cite{Harvey:1998bx,Yi:2001bz,Hu:2011ji} , as well as in pure 6D (2,0) superconformal theory \cite{Henningson:1998gx,Bastianelli:2000hi,Maxfield:2012aw}.
 In the AdS/CFT of holographic hydrodynamics \cite{Herzog:2002fn,Herzog:2003ke} (see \cite{Rangamani:2009xk} for a review), inclusion of M5 branes led to the $n^3$ behavior of pressure and shear viscosity of 6D strongly-coupled supersymmetric plasma at non-zero temperature \cite{Herzog:2002fn,Herzog:2003ke,Natsuume:2008iy}.

The close relation between 6D effective worldvolume theory of M5 branes and 5D effective worldvolume description of D4 branes goes back to the equivalence of M-theory and the strongly-coupled type IIA string theory. This duality allows one to study M5 branes through D4 brane/open string techniques. Although the $n^3$ entropy growth was originally discovered in the supergravity setup \cite{Klebanov:1996un}, the role of open strings in closed string theory has been widely recognized. Different aspects of open string feedbacks to closed string/supergravity models include, for instance, the seminal KLT relation \cite{Kawai:1985xq} between open and closed string tree amplitudes  (see \cite{Bern:2010yg,BjerrumBohr:2010yc} for recent progress), as well as the CSW \cite{Cachazo:2004kj} and BCFW \cite{Britto:2005fq} recursive relations. The role of open string loops in D-brane
curvature generation was proposed in \cite{Park:2011if,Park:2011bg} (and refs therein).
Recently, the $n^3$ entropy growth was obtained \cite{Hatefi:2012sy} in a D0-based SYM setup by including Myers' terms \cite{Myers:1999ps,Garousi:2008ge,Hatefi:2010ik,Hatefi:2012ve} and subsequently applying the localization technique of \cite{Witten:1988ze, Kapustin:2009kz}.

 One of the goals of this work is to search for additional examples of black M-brane configurations that display the characteristic $S\sim n^3 T^5$ near-extremal entropy-temperature relation of black M5 branes. Configurations of KKW/M2/M5 black branes are natural places for searching for the systems with the $n^3$ behavior: all of these constituents (KKW,M2,M5) enter M-algebra \cite{Sorokin:1997ps} in a democratic way. Below we observe the
 characteristic $n^3$ growth in a system of KKW/M2/M5 black branes, a generalization of the M2/M5 inter\-po\-la\-ting solution of \cite{Izquierdo:1995ms,Green:1996vh}, and formulate conditions under which the behavior arises. One novelty of the solution worth mentioning is that it has a single harmonic function in M2/M5 sector and certain ``duality" in thermodynamics under the interchange of the roles of KKW and M5.
 We will comment on its implication for our recent work \cite{Hatefi:2012sy} in the conclusion.

 As we will discuss, the pertaining configurations of lower dimensional branes, lying inside or intersecting with M5 branes, is thermodynamically equivalent to the stack of coincident M5's. It should be a manifestation of the Myers' effect \cite{Emparan:1997rt,Myers:1999ps} on the supergravity side, and provides a rationale for the setup of our work \cite{Hatefi:2012sy}. Recall that in the supergravity context, the dielectric effect should manifest itself as the dissolution of lower dimensional branes into higher dimensional ones. The dielectric effect was observed in other supergravity setups that correspond to the bound states of D0/D4 branes \cite{Marolf:2001np}, flux branes \cite{Costa:2001ifa,Brecher:2001xj}, and tubular branes \cite{Emparan:2001rp,Emparan:2001ux}.

The same $n^3$ scaling is found below for descents of our KKW/M2/M5 solution to lower dimensions. Examples of non-intersecting black branes with the $n^3$ entropy include D3-branes in 9D, D2 branes in 8D, D1-branes in 7D and D0 branes in 6D. We explicitly check their near-extremal entropy growths, and confirm their $n^3$ behaviors.

\ssk
The rest of the paper is organized as follows. In Section 2 we review thermodynamics of non-extremal M5 branes and the equation of state of coincident M5 black branes in the near-extremal limit. We establish the $n^3$ entropy growth, and find the shear viscosity/entropy bound of M5 branes within the AdS/CFT hydrodynamical limit. The same section contains a near-extremal entropy analysis of  non-extremal M5/KKW and M5/M2/KKW intersecting branes. The analysis leads to formulation of a criterion for constructing supergravity solutions that have the same equation of state as that of M5 branes. In Section 3 we construct the KKW/M2/M5 interpolating solution with the $n^3$ near-extremal entropy growth. Dimensional reduction to a type IIA configuration is considered in Section 4. There, we also construct the descents of the IIA D4 black-brane solution, and examine their thermodynamical properties. Section 5 contains discussion of the results and comments on further studies. For self-containedness of the paper, we have collected a brief summary of thermodynamical characteristics of black branes and the basics of toroidal dimensional reduction in two Appendices.



\ssk

\ssk

\section{Black M branes and their intersections}

In this section we consider various configurations of black branes of M-algebra \cite{Sezgin:1996cj,Sorokin:1997ps} -- KK-waves, M2 and M5 branes -- and their near-extremal thermodynamics.
Some of these configurations exhibit the $n^3$ scaling while others do not.

We start with a brief review of \cite{Klebanov:1996un} followed by comments on the hydrodynamics at the end.
A stack of $n$ coincident black M5 branes is described by the following magnetic-type solution \cite{Gueven:1992hh,Horowitz:1991cd,Duff:1996hp} 
\be
ds^2_{11}=H_5^{-1/3}(r)\left[-f(r)dt^2+dx_1^2+\dots+dx_5^2 \right]+H_5^{2/3}(r)\left(\fr{dr^2}{f(r)}+r^2 d\Omega^2_4 \right),\la{M5Bng}
\ee
\be
H_5(r)=1+n\fr{h_0^3}{r^3},\qquad f(r)=1-\fr{\m^3}{r^3}, \la{M5BnH}
\ee
\be
F_{[4]}=3nh_0^3 \left(1+\fr{\m^3}{n h_0^3}\right)^{1/2}\e_4,\qquad F_{[4]}=dC_{[3]}
\la{F4M5Bn}
\ee
to 11D supergravity equations of motion \cite{Cremmer:1978km}.
$\e_4$ is the volume form of the unit 4-sphere; $h_0$ is associated with the M5 charge.
Calculating the entropy and temperature of the system by standard methods (cf. eqs. \rf{qpS}, \rf{qpT}),
one gets
\be
S_5=\fr{l^5 \Omega_4}{4G_{11}}\left(1+n\fr{h_0^3}{\m^3}\right) ^{1/2} \m^4 ,
\la{SnM5B}
\ee
\be
T_5=\fr3{4\pi} \left(1+n\fr{h_0^3}{\m^3}\right) ^{-1/2}\m^{-1}.
\la{TnM5B}
\ee
Another important characteristic of the system is its ADM mass.
Adopting the general expression \rf{qpADM}, we get
\be
M_5=\fr{l^5\Omega_4}{16 \pi G_{11}}\left[ 4 \mu^3+3nh_0^3 \right].
\la{M5ADM}
\ee
Taking the extremal limit, $\m \ra 0$, results in the following mass-charge bound relation
\be
M^\0=\fr{l^5\Omega_4}{16 \pi G_{11}} 3nh_0^3 .
\la{M5M0}
\ee
(The entropy and the temperature in the {\em extremal} limit vanish as they should.)
To compute $S(T)$ dependence in the near-extremal limit, we perturb, following \cite{Klebanov:1996un}, the ADM mass around its bound value \rf{M5M0},
\be
M_5=M_5^\0 \left(1+\fr{\d M}{M^\0}\right),\qquad \d M \ll M^{\0}.
\la{M5ADMexpand}
\ee
Therefore, $\m$ satisfies $\m \ll 1$ in the near-extremal limit. Roughly, $\d M$ scales as $\d M \sim \m^3$; more precisely, one gets
\be
\m=\left(\fr{\d M}{4} \right)^{1/3}\a_5^{1/3},\qquad \a_5=\fr{16\pi G_{11}}{l^5 \Omega_4}.
\la{muNEM5}
\ee
Replacement of $\mu$ in \rf{SnM5B} by its near-extremal value \rf{muNEM5} leads to
\be
S_5=\fr{l^5 \Omega_4}{4G_{11}}\left(1+n\fr{h_0^3}{\m^3}\right) ^{1/2}\m^4\approx \fr{l^5 \Omega_4}{4G_{11}}h_0^{3/2}n^{1/2}\m^{5/2}\sim n^{1/2}(\d M)^{5/6}.
\la{SM5NE}
\ee
The Hawking temperature \rf{TnM5B} takes
\be
T_5\approx \fr3{4\pi} h^{-3/2}_0 n^{-1/2} \m^{1/2} \sim n^{-1/2} (\d M)^{1/6}.
\la{TH5NE}
\ee
Interpreting $\d M$ as the energy $E$ (cf. \cite{Klebanov:1996un}), we arrive at
\be
S_5 \sim n^{1/2} E^{5/6},\qquad T_5\sim n^{-1/2} E^{1/6} ,
\la{STM5}
\ee
in agreement with the first law of thermodynamics.
From the latter expressions we get the Klebanov-Tseytlin result
\be
S_{M5} \sim n^3 T^5,
\la{SM5nT}
\ee
the near-extremal $n^3$ entropy growth.

\ssk
As observed in \cite{Klebanov:1996un}, the near-extremal entropy of black M5 branes may be recovered in terms of a weakly-interacting ideal gas of 6D massless particles. Indeed, the explicit value of near-extremal entropy density (per unit volume) of black M5 branes \cite{Klebanov:1996un,Herzog:2002fn} is given by
\be
S_5=2^7 3^{-6}\p^3 n^3 T^5 ,
\la{KTS5}
\ee
and is the same as the entropy of a gas of $N_5=2^{10}3^{-6} 5n^3$ massless bosons and fermions, whose degrees of freedom match  due to the supersymmetry of M5 branes.

In the context of  AdS/CFT hydrodynamics \cite{Policastro:2002se,Policastro:2002tn}, one of the test stone quantities is the ratio of the shear viscosity \cite{Herzog:2002fn,Herzog:2003ke,Natsuume:2008iy}
\be
\eta_5=\fr{\pi^2}2 \left(\fr23 \right)^6 n^3 T^5
\la{SV5}
\ee
to the entropy \rf{KTS5},
\be
\eta_5/S_5=\fr1{4 \pi}.
\la{etaSbound}
\ee
It precisely corresponds to the bound value of the universal $\eta /S$ relation
\be
\eta /S \ge 1/4\pi,
\la{etaSuni}
\ee
that was conjectured in \cite{Kovtun:2004de}.\footnote{This bound was violated, e.g., in higher derivative gravity \cite{Brigante:2007nu}, or in anisotropic plasma \cite{Rebhan:2011vd}.}

\subsection{Various other configurations containing M5's}


With the above review we pose the following question: could the characteristic $n^3$ entropy growth be observed in other coincident black brane systems?
After examining various other configurations containing M5 branes below, we discuss the necessary
conditions in section 2.2.

\subsubsection{M5/KKW bound state in the near-extremal limit}

To find the answer for the question above, let us consider a configuration of M5 branes
with added M-waves (KKWs). The procedure is well-known \cite{Tseytlin:1996bh,Tseytlin:1996hi,Costa:1996re,Ohta:1997wp},
and the non-extremal KKW/M5 bound state solution has the following form
\[
ds^2_{11}=H_5^{2/3}\left[H_5^{-1}(-K^{-1}fdt^2+Kd\hat{x}^2_1)+H^{-1}_5 (dx^2_2+\dots+dx^2_5) \right.
\]
\be
\left . +f^{-1}dr^2+r^2 d\Omega^2_4 \right] .
\la{WM5}
\ee
where
\be
H_5(r)=1+n_5\fr{h_5^3}{r^3},~ K(r)=1+n_0\fr{k_0^3}{r^3},~ f(r)=1-\fr{\m^3}{r^3},~ d\hat{x}^2_1=[dx_1+(K^{-1}-1)dt]^2 ,
\la{WM5KHf}
\ee
\be
F_{[4]}=3n_5h_5^3 \left(1+\fr{\m^3}{n_5 h_5^3}\right)^{1/2}\e_4 .
\la{WM5F4Bn}
\ee
The bound state is formed by $n_0$ black KK-waves that travel along the $x_1$ direction of the world-volume of $n_5$ coincident black M5 branes.

Computing the ADM mass, entropy and  temperature of the non-extremal configuration, one gets\footnote{Even though the metric \rf{WM5} contains $g_{x_1 t}$ non-diagonal term, it is diagonal in the $\hat{x}_1$ coordinate; one may still use \rf{qpS}, \rf{qpT}, \rf{qpADM}. There are two reasons behind this: the equality of the areas of black KKW/M5 bound-state configurations and the equality of the corresponding time-like Killing vectors in $x_1$ and $\hat{x}_1$ coordinates. The former leads to the same entropy for both coordinates and the latter to the same values of temperature.}
\be
M=\fr{l^5 \Omega_4}{16\pi G_{11}}\left[ 4\m^3+3n_0k_0^3+3n_5 h^3_5 \right],
\la{WM5M}
\ee
\be
S=\fr{l^5 \Omega_4}{4 G_{11}}\left(1+n_5 \fr{h^3_5}{\m^3}\right)^{1/2}\left(1+n_0 \fr{k^3_0}{\m^3}\right)^{1/2} \m^4,
\la{WM5S}
\ee
\be
T=\fr3{4\pi}\left(1+n_5 \fr{h^3_5}{\m^3}\right)^{-1/2}\left(1+n_0 \fr{k^3_0}{\m^3}\right)^{-1/2} \m^{-1}.
\la{WM5T}
\ee
In the near-extremal limit $\d M\sim \m^3 \ll 1$,
\be
S\sim \sqrt{n_0 n_5}(\d M)^{1/3},\qquad T\sim (\sqrt{n_0 n_5})^{-1} (\d M)^{2/3}.
\ee
Therefore, the entropy of black KKW/M5 branes obeys
\be
S\sim (n_0 n_5)^{3/4} T^{1/2} .
\la{WM5ST}
\ee
At first sight, the $S(T)$ dependence of \rf{WM5ST} is far from the $n^3$ dependence of \rf{SM5nT}.
However, once $n_0$  in \rf{WM5M}--\rf{WM5T} is set to zero, \rf{WM5M}--\rf{WM5T} turn into \rf{M5ADM}, \rf{SnM5B}, \rf{TnM5B}; the near-extremal entropy will have the $n^3$ growth. The same thing happens when the number of KK-waves is restricted to $n_0 k_0^3 \ll 1$.\footnote{Note, however, that to keep the supergravity approximation, $n_0 \gg 1$ is still needed, and $n_5 h_5^3 \gg 1$.}
Tuning the parameters in this way, the horizon of the configuration is formed solely by M5 branes. It follows that the thermodynamics of the KKW/M5 solution in the $n_0 k_0^3 \ll 1$ limit is completely determined by M5 black brane characteristics. In other words, the near-extremal entropy grows as
\be
S_{KKW/M5}\sim n_5^3 T^5, \qquad (n_0 k_0^3 \ll 1, ~ n_0 \gg 1) .
\la{KKWM5Sne}
\ee


 One thing worth pointing out is the fact that the respective ``charges'' $h_5$ and $k_0$ of the M5 branes and KKWs
 enter \rf{WM5M}--\rf{WM5T} in a symmetric way. Expressions for $M,S,T$ are invariant under the interchange of $n_0 \leftrightarrow n_5$, $h_5 \leftrightarrow k_0$. This symmetry suggests the following interpretation upon considering the interchange of the roles of M5's and KKWs. The bound state is formed with a number of KK-waves in the $n_0 k_0^3 \gg 1$
 limit and a number of M5 branes in the $n_5 h_5^3 \ll 1$, $n_5 \gg 1$ limit. With this choice, the thermodynamics of KK-waves models that of the black M5 branes; the near-extremal entropy becomes
\be
S_{KKW/M5}\sim n_0^3 T^5, \qquad (n_5 h_5^3 \ll 1, ~ n_5 \gg 1),
\la{KKWM5Sne1}
\ee
It may be interpreted as a manifestation of the Myers' effect \cite{Myers:1999ps} of polarization of KK-waves (M0 branes) into M5 branes.

\subsubsection{M2/M5 intersection}

 Let us add M2 branes into the setup, and ``blacken'' the extremal solutions of M2/M5 and KKW/M2/M5 brane intersections respectively. The configuration that we consider in this subsection will not lead to $n^3$
behavior.
Using \cite{Argurio:1997gt}, the number of common dimensions $\bar{q}$ over which M$q$ and M$p$ branes intersect is evaluated as
\be
\bar{q}=\fr{(q+1)(p+1)}{9}-1 ,\qquad \bar{q} \in \mathbb{Z}, \qquad \bar{q}>0.
\la{interq}
\ee
%
%
\ni Let us turn to the single M2/M5 branes intersection over a line ($\bar{q}=[(5+1)(2+1)/9]-1=1$), and set the line along the $x_1$ direction. The corresponding intersection diagram looks as follows

\ssk
\ssk
\begin{tabular}{cccccccccccc}
& $t$ & $x_1$ & $x_2$ & $x_3$ & $x_4$ & $x_5$ & $x_6$ &  $x_7$ & $x_8$ & $x_9$ & $x_{10}$\\
$\mathrm{M2}$ & $\times$ &$\times$  & & & & & $\times$ & & & & \\
$\mathrm{M5}$ & $\times$ & $\times$&$\times$ & $\times$ & $\times$ & $\times$ &&&&&
\end{tabular}

\ssk \ssk
\noindent The line element\footnote{Since we are mostly interested in computing $M,S$ and $T$, we skip the details on the gauge sector of the solution here and below.} is \cite{Tseytlin:1996bh,Tseytlin:1996hi}
\[
ds^2_{11}=H_2^{1/3}H_5^{2/3}\left[(H_2H_5)^{-1}(-dt^2+dx_1^2)+H_2^{-1}dx_6^2 \right.
\]
\be
\left. +H_5^{-1}(dx_2^2+dx_3^2+\dots+dx_5^2)+dr^2+r^2 d\Omega^2_3 \right]  ,
\la{M2M5g}
\ee
where
\bea
r^2&=&x_7^2+x_8^2+x_9^2+x_{10}^2 \, ,  \nn\\
H_i(r)&=&1+\fr{h_i^2}{r^2},\qquad i=2,5
\la{M2M5H}
\eea
are harmonic functions in the 4D space that is spanned by  the radial coordinate $r$ and three angles of $d\Omega_3$.
The generalization of \rf{M2M5g}, \rf{M2M5H} to the non-extremal multiple intersecting M2/M5 branes takes
\[
ds^2_{11}=H^{1/3}_2 H^{2/3}_5 \left[ (H_2H_5)^{-1}(-fdt^2+dx^2_1)+H_2^{-1}dx^2_6 \right.
\]
\be
\left. +H^{-1}_5 (dx^2_2+dx^2_3+\dots+dx^2_5)+f^{-1}dr^2+r^2 d\Omega^2_3 \right],
\la{M2M5Bds}
\ee
\be
H_2(r)=1+n_2\fr{h_2^2}{r^2},\qquad H_5(r)=1+n_5 \fr{h_5^2}{r^2},\qquad f(r)=1-\fr{\m^2}{r^2} \, ,
\la{M2M5BHf}
\ee
where $n_2$ and $n_5$ are the M2 and M5 brane numbers respectively.
The ADM mass, the entropy and the temperature of the system are
\be
M=\fr{l^6 \Omega_3}{16\pi G_{11}}\left[ 3\m^2+2n_2 h_2^2+2n_5 h_5^2 \right],
\la{M2M5M}
\ee
\be
S=\fr{l^6 \Omega_3}{4 G_{11}}\left(1+n_2 \fr{h^2_2}{\m^2}\right)^{1/2}\left(1+n_5 \fr{h_5^2}{\m^2}\right)^{1/2} \m^3,
\la{M2M5S}
\ee
\be
T=\fr2{4\pi}\left(1+n_2 \fr{h^2_2}{\m^2}\right)^{-1/2}\left(1+n_5 \fr{h_5^2}{\m^2}\right)^{-1/2} \m^{-1} \, .
\la{M2M5T}
\ee
They take
\be
S\sim n_2^{1/2}n_5^{1/2} (\d M)^{1/2},\quad T\sim n_2^{-1/2}n_5^{-1/2} (\d M)^{1/2} .
\ee
in the near-extremal limit ($\m\sim (\d M)^{1/2}$ for the present case).
This result implies
\be
S\sim n_2 n_5 T .
\la{M2M5ST}
\ee
We conclude that it is impossible to recover the $n^3$ entropy growth in the current case.

\subsubsection{boosted non-extremal M2/M5 intersection}

Finally, let us check the near-extremal entropy growth in the system of non-extremal KKW/M2/M5 intersecting branes.
The solution that corresponds to KKW/M2/M5 configuration can be obtained from \rf{M2M5Bds} by inclusion of M5 branes boosted along the intersecting direction (it is the $x_1$-direction in our case) \cite{Tseytlin:1996bh,Costa:1996re,Ohta:1997wp}. The boosted form of \rf{M2M5Bds} is
\[
ds^2_{11}=H^{1/3}_2 H^{2/3}_5 \left[ (H_2H_5)^{-1}(-K^{-1}fdt^2+Kd\hat{x}^2_1)+H_2^{-1}dx^2_6 \right.
\]
\be
\left. +H^{-1}_5 (dx^2_2+dx^2_3+\dots+dx^2_5)+f^{-1}dr^2+r^2 d\Omega^2_3 \right],
\la{WM2M5Bds}
\ee
\be
H_2(r)=1+n_2\fr{h_2^2}{r^2},\qquad H_5(r)=1+n_5 \fr{h_5^2}{r^2},\qquad f(r)=1-\fr{\m^2}{r^2} ,
\la{WM2M5BHf}
\ee
\be
K(r)=1+n_0 \fr{k^2_0}{r^2},\qquad d\hat{x}^2_1=\left[dx_1+(K^{-1}-1)dt\right]^2 \, ,
\la{WM2M5Kx1}
\ee
where $n_0$, $n_2$ and $n_5$ count the number of KK waves, M2 and M5 branes respectively.
After some algebra, one can obtain
\be
M=\fr{l^6 \Omega_3}{16\pi G_{11}}\left[ 3\m^2+2n_2 h_2^2+2n_5 h_5^2 +2n_0k^2_0\right],
\la{WM2M5M}
\ee
\be
S=\fr{l^6 \Omega_3}{4 G_{11}}\left(1+n_2 \fr{h^2_2}{\m^2}\right)^{1/2}\left(1+n_5 \fr{h_5^2}{\m^2}\right)^{1/2} \left(1+n_0\fr{k_0^2}{\m^2}\right)^{1/2}\m^3,
\la{WM2M5S}
\ee
\be
T=\fr2{4\pi}\left(1+n_2 \fr{h^2_2}{\m^2}\right)^{-1/2}\left(1+n_5 \fr{h_5^2}{\m^2}\right)^{-1/2} \left(1+n_0\fr{k_0^2}{\m^2}\right)^{-1/2} \m^{-1} .
\la{WM2M5T}
\ee
In the near-extremal limit of $(\d M)\sim \m^2$, the entropy and temperature scale
according to
\be
S\sim n_2^{1/2}n_5^{1/2} n_0^{1/2} (\d M)^0,\qquad T\sim n_2^{-1/2}n_5^{-1/2} n_0^{-1/2} (\d M).
\la{STNE}
\ee
In this case, one encounters the near-extremal entropy independence of energy.
 Also, as in the previous subsection, the desired $n^3$ entropy does not result in this setup.


\subsection{Condition for reproduction of $n^3$ scaling}

Let us pause and summarize the results so far. With appropriately chosen limits of the parameters (see \rf{KKWM5Sne} and \rf{KKWM5Sne1}), we could observe the $n^3$ entropy scaling in the boosted M5 solution \rf{WM5} above. With a different set of scalings, the thermodynamics of $n$ coincident black M5 branes could be completely modeled by the KK-waves that propagate inside the stack of M5's.
We have interpreted this phenomenon as a manifestation of the Myers' effect on the supergravity side.
Meanwhile, we have concluded that the $n^3$ near-extremal entropy growth cannot be reached with the M2/M5 (eq.\rf{M2M5Bds}) or the KKW/M2/M5 intersecting systems (eq.\rf{WM2M5Bds}) of delocalized M black branes.\footnote{In this paper we only focus on the delocalized solutions. The localized versions of the systems considered here (see, e.g., \cite{Edelstein:1998vs,Yang:1999ze,Aref'eva:1998uh,Youm:1999ti}) are not the subject of the paper. } We would like to elaborate on
this further here.

It is well-known that thermodynamics of black branes configurations is completely encoded in the structures of the harmonic functions and the ``blackening'' factor(s) that appear in the solution. (One can see this just by considering \rf{qpds}, \rf{qpS}, \rf{qpT} and \rf{qpADM}). To reproduce the same thermodynamics of a given system (say, the M5 brane system in the present case) by another system (say, the boosted M5 branes, or their intersections with KK waves and M2 branes), it is necessary to assure that the two configurations under consideration have {\it the same} entropy/temperature dependence on the non-extremality parameter $\m$ in the blackening factor $f(r)$. Comparing \rf{WM5S} and \rf{WM5T} with \rf{SnM5B} and \rf{TnM5B}, one can see that it is indeed the case: the entropy and the temperature of two systems have the same dependence on $\m$. Meanwhile, the entropy of M2/M5 branes given in \rf{M2M5S} shows different dependence on $\m$ compared with \rf{SnM5B}. It leads to a completely different entropy-temperature equation of state in the end, and, consequently, to different thermodynamics. Although
in the present paper we do not consider solutions of partially localized branes, this observation remains true for the latter case as well.

Therefore, it is clear that compelling candidates for intersecting solutions of M branes that could potentially reproduce (exactly or within some special choice of the parameters) the $n^3$ growth of the black M5 branes are those whose harmonic functions have $1/r^3$ dependence. Put differently, $d \Omega_4$ volume element must be present in the solution.

In what follows we will focus on the M2/M5 interpolating solution of \cite{Izquierdo:1995ms,Green:1996vh} and its boosted non-extremal version \cite{Costa:1996re}. They both fall into the criterion just drawn. Afterwards, we will show that solutions obtained by dimensional reduction of M interpolating branes and double dimensional reduction of M5 branes exhibit the $n^3$ near-extremal entropy behavior.


\section{M-branes interpolating solutions}

\setcounter{equation}{0}

Consider the following solution that interpolates between M2 and M5 branes \cite{Izquierdo:1995ms,Green:1996vh,Costa:1996re}
\[
ds^2_{11}=(H\Ht)^{1/3}\left[H^{-1}(-fdt^2+dx^2_1+dx^2_2)+\Ht^{-1}(dx^2_3+dx^2_4+dx^2_5) \right.
\]
\be
\left. +f^{-1}dr^ 2+r^2d \Omega^2_4\right] ,
\la{M2M5int}
\ee
\be
\hat{F}_4=\fr12\cos\zeta \ast dH+\fr12\sin \zeta \,dH^{-1}\e_3+\fr32\sin2\zeta \,H^{-2} dH \, \bar{\e}_3 \, ,
\la{M2M5F4int}
\ee
\be
H(r)=1+n\fr{h^3}{r^3},\qquad \Ht=\sin^2 \zeta+H\cos^2 \zeta,\qquad  f(r)=1-\fr{\m^3}{r^3} ,
\la{HHtfint}
\ee
where $\e_3$ and $\bar{\e}_3$ are volume forms on $\mathbb{M}^3$ and $\mathbb{E}^3$ parameterized respectively by $(t,x_1,x_2)$ and $(x_3,x_4,x_5)$; $\ast$ is the Hodge dual of $\mathbb{E}^5$ that is transverse to the M5 branes.
This solution is completely different from the standard intersection of M2/M5  \rf{M2M5Bds}. In contrast to \rf{M2M5Bds} whose intersection diagram was given in Sec. 2.3,
the solution \rf{M2M5int}--\rf{HHtfint} describes M2 branes entirely lying within M5 branes:

\ssk
\ssk
\begin{tabular}{cccccccccccc}
& $t$ & $x_1$ & $x_2$ & $x_3$ & $x_4$ & $x_5$ & $x_6$ &  $x_7$ & $x_8$ & $x_9$ & $x_{10}$\\
$\mathrm{M2}$ & $\times$ &$\times$  & $\times$& & & &  & & & & \\
$\mathrm{M5}$ & $\times$ & $\times$&$\times$ & $\times$ & $\times$ & $\times$ &&&&&
\end{tabular}

\ssk
\ssk \noindent
Unlike \rf{M2M5Bds} which preserves 1/4 of the original supersymmetry, \rf{M2M5int} preserves 1/2. It is constructed using a {\em single} harmonic function $H(r)$; $\sin \zeta=0$ corresponds to the black M5 brane solution \rf{M5Bng}--\rf{F4M5Bn} and
$\cos\zeta=0$ gives a configuration of black M2 branes.
As we will see shortly, dimensional reduction of \rf{M2M5int} leads to type IIA interpolating solutions of D2 branes within D4 branes, or F1s within D4s when reduced along any of $(x_3,x_4,x_5)$.


The thermodynamics pertaining to \rf{M2M5int} are
\be
M=\fr{l^5 \Omega_4}{16 \pi G_{11}}\left[ 4\m^3+3n h^3 \right],
\la{M2M5intM}
\ee
\be
S=\fr{l^5 \Omega_4}{4G_{11}} \left(1+n\fr{h^3}{\m^3}\right )^{1/2} \m^4 ,
\la{M2M5intS}
\ee
\be
T=\fr3{4\pi}\left(1+n\fr{h^3}{\m^3}\right)^{-1/2} \m^{-1}.
\la{M2M5intT}
\ee
These expressions are precisely those of the stack of black M5 branes (cf. \rf{SnM5B}--\rf{M5ADM}). One can easily check that
\be
S\sim n^3 T^5
\la{M2M5intST}
\ee
in the near-extremal limit. The solution describes a configuration of the M2 branes
completely dissolved in the M5 branes.


The boosted version of \rf{M2M5int} involving the KK-wave along, say, the $x_1$ direction is described by the following ansatz
\[
ds^2_{11}=(H\Ht)^{1/3}\left[H^{-1}(-K^{-1}fdt^2+Kd{\hat x}^2_1+dx^2_2) \right .
\]
\be
\left . +\Ht^{-1}(dx^2_3+dx^2_4+dx^2_5)+f^{-1}dr^ 2+r^2d \Omega^2_4\right] ,
\la{WM2M5int}
\ee
with
\[
H(r)=1+n\fr{h_0^3}{r^3},\quad \Ht=\sin^2 \zeta+H\cos^2 \zeta,\quad  f(r)=1-\fr{\m^3}{r^3},
\]
\be
K(r)=1+n_0 \fr{k_0^3}{r^3},\quad d{\hat x}_1^2=[dx_1+(K^{-1}-1)dt]^2 .
\la{WHHtfint}
\ee
The gauge part of the solution coincides with \rf{M2M5F4int}.
The solution \rf{WM2M5int} with the harmonic functions and the blackening factor of \rf{WHHtfint} is a generalization of the boosted solution of \cite{Costa:1996re}. Its thermodynamics is quite similar to the boosted M5 branes solution considered in Sec. 2.1.1. Therefore, tuning the parameters of  \rf{WHHtfint} to $n h_0^3 \ll 1$, $n \gg 1$, $n_0 k_0^3 \gg 1$,  the KKW/M2/M5 interpolating solution becomes thermodynamically equivalent to the stack of $n_0$ black M5 branes.\footnote{The fact that only a single harmonic function appears together with this ``duality'' in thermodynamics may have an interesting implication
related to our recent work \cite{Hatefi:2012sy}. We will comment on it in the conclusion.} We interpret  this phenomenon as a realization of the Myers' effect of polarization of branes.

\section{Dimensional reduction of black M-branes solutions to D=10 IIA ones}

\setcounter{equation}{0}

 Let us consider the dimensionally reduced versions of the previously obtained solutions to 11D supergravity. Basics of the toroidal dimensional reduction are reviewed in Appendix B.



\ssk
\ssk
The $n^3$ near-extremal entropy growth can be found in the 10D type IIA $n$
D4 black branes that is obtained by dimensional reduction of \rf{M5Bng}.\footnote{This was noticed early in \cite{Ryang:1999bh}.} Since general expressions for the ADM mass, the entropy and the temperature are typically written in the Einstein frame, it is convenient to reduce \rf{M5Bng} by use of the Einstein frame reduction ansatz
\be
{g}^{(11)}_{MN}\, dx^M dx^N=e^{-\phi/6}g^{(10)}_{mn} \, dx^m  dx^n+e^{4\phi/3}\left(dx^{\flat}+A_m dx^m \right)^2 ,
\la{KKansE}
\ee
where $x^{\flat}$ denotes the direction of the reduction.
Reducing \rf{M5Bng} along one of $x_{1,2,3,4,5}$ (say, $x^\flat=x_5$) one gets
\be
ds^2_{10}=H_4^{-3/8}\left[-f(r)dt^2+dx_1^2+\dots +dx^2_4 \right]+H_4^{5/8}\left(\fr{dr^2}{f(r)}+r^2 d\Omega^2_4 \right),
\la{D4E}
\ee
\be
H_4(r)=1+n\fr{h_0^3}{r^3},\qquad f(r)=1-\fr{\m^3}{r^3}, \qquad e^\phi=H_4^{-1/4}.
\la{D4HfE}
\ee
The $C_{[3]}$ gauge field (other gauge fields are set to zero) is given by
\be
F_{[4]}=3nh_0^3\left(1+\fr{\m^3}{nh^3_0}\right)^{1/2}\e_4 \, .
\la{D4F4}
\ee
The solution \rf{D4E}--\rf{D4F4} describes the background of $n$-coincident black D4 branes of type IIA theory (see, e.g., \cite{ArgurioPhD}).

Computing the ADM mass, entropy and temperature, one gets
\be
M_4=\fr{l^4 \Omega_4}{16\pi G_{10}}\left[ 4\m^3+3nh_0^3 \right] ,
\la{D4M}
\ee
\be
S_4=\fr{l^4 \Omega_4}{4G_{10}} \left(1+n\fr{h_0^3}{\m^3}\right)^{1/2} \m^4 ,
\la{D4S}
\ee
\be
T_4=\fr3{4\pi} \left(1+n\fr{h_0^3}{\m^3}\right)^{-1/2} \m^{-1} .
\la{D4T}
\ee
Comparing the latter expressions with \rf{SnM5B}--\rf{M5ADM} it is easy to notice, after identifying $G_{11}=l G_{10}$, the similarity between \rf{D4M}--\rf{D4T} and the corresponding quantities of $n$ coincident black M5 branes. Hence, one may conclude that the thermodynamics of $n$ coincident black D4 branes leads to the entropy-temperature dependence
\be
S_{D4}\sim n^3 T^5
\la{STD4}
\ee
in the near-extremal limit.

The solution \rf{WM5}, once dimensionally reduced along the KK waves propagation direction,  leads to the D0/D4 bound state. Indeed, by using the reduction ansatz \rf{KKansE} and reducing along the $x^\flat=x_1$, we arrive at\footnote{See, e.g., \cite{Tseytlin:1996bh,Tseytlin:1996hi,Argurio:1997gt,Ohta:1997gw} for a general discussion on D$q$/D$p$ intersecting branes.}
\be
ds^2_{10}=H_4^{5/8}K_0^{1/8}\left[-(H_4K_0)^{-1} fdt^2+H_4^{-1}(dx^2_2+\dots+dx^2_5)+f^{-1}dr^2+r^2d\Omega^2_4\right],
\la{dsD0D4B}
\ee
\be
A_{[1]}=(K_0^{-1}-1)dt,\qquad e^{4\phi/3}=H_4^{-1/3}K_0,
\la{AphiD0D4B}
\ee
\be
H_4(r)=1+n_4\fr{h_4^3}{r^3},\quad K_0(r)=1+n_0\fr{k_0^3}{r^3},\quad f(r)=1-\fr{\m^3}{r^3} \, .
\la{D0D4KHf}
\ee
The near-extremal $S(T)$ dependence of the D0/D4 brane intersection
\be
S\sim (n_0 n_4)^{3/4} T^{1/2}
\la{D0D4ST}
\ee
matches with \rf{WM5ST}. Therefore, the results of Sec. 2.1.1 are directly applied to the D0/D4 case, and one may scale the parameters appropriately so that that thermodynamics of black D0 branes will reproduce that of D4 branes in the near-extremal limit,
\be
S_{D0/D4}\sim n_0^3 T^5 \quad (n_0 k_0^3 \gg 1,~ n_4 h^3_4 \ll 1,~ n_4 \gg 1) .
\la{D0D4Sne}
\ee
Again, it should be a manifestation of the Myers' effect of polarization of D0 branes to D4 branes on supergravity side.

\subsection{D0/D2/D4 interpolating solution}

\ssk
Dimensional reduction of \rf{WM2M5int} along any of $x_{3,4,5}$ directions leads to the D0/D2/D4 interpolating solution, D2/D4 part of which (in the string frame) was constructed in \cite{Green:1996vh}.
We will reduce along $x_3$ coordinate, which is associated with the traveling direction of the KK-waves. The eleven dimensional metric is
\[
ds^2_{11}=(H\Ht)^{1/3}\left[H^{-1}(-K^{-1}fdt^2+dx^2_1+dx^2_2)+\Ht^{-1}(Kd{\hat x}^2_3+dx^2_4+dx^2_5) \right.
\]
\be
\left . +f^{-1}dr^ 2+r^2d \Omega^2_4\right] ,
\la{WM2M5int3}
\ee
where
\[
H(r)=1+n\fr{h_0^3}{r^3},\quad \Ht=\sin^2 \zeta+H\cos^2 \zeta,\quad  f(r)=1-\fr{\m^3}{r^3},
\]
\be
K(r)=1+n_0 \fr{k_0^3}{r^3},\quad d{\hat x}_3^2=[dx_3+(K^{-1}-1)dt]^2 .
\la{WHHtfint3}
\ee
Reducing along $x_3$, one gets
\[
ds^2_{10}=H^{3/8}\Ht^{1/4}K^{-3/8}\left[H^{-1}(-K^{-1}fdt^2+dx_1^2+dx^2_2)+\Ht^{-1}(dx_4^2+dx_5^2) \right.
\]
\be
\left . +f^{-1}dr^ 2+r^2d \Omega^2_4\right] ,
\la{D0D2D4int}
\ee
\be
e^\phi=H^{1/4}\Ht^{1/2}K^{-3/4},\qquad A_{[1]}=(K^{-1}-1)dt .
\la{D0D2D4phi}
\ee
in the Einstein frame.
In the $K=1$ setting, \rf{D0D2D4int}, \rf{D0D2D4phi} reduce to the D2/D4 interpolating solution of \cite{Green:1996vh}. One can show that
\be
M=\fr{l^4 \Omega_4}{16\pi G_{10}}\left[ 4\m^3+3nh_0^3 \right] ,
\la{D2D4intM}
\ee
\be
S=\fr{l^4 \Omega_4}{4G_{10}} \left(1+n\fr{h_0^3}{\m^3}\right)^{1/2} \m^4 ,
\la{D2D4intS}
\ee
\be
T=\fr3{4\pi} \left(1+n\fr{h_0^3}{\m^3}\right)^{-1/2} \m^{-1} .
\la{D2D4intT}
\ee
These expressions coincide with those of the interpolating solution of M2/M5 (compare \rf{D2D4intM}--\rf{D2D4intT} with \rf{M2M5intM}--\rf{M2M5intT}, identifying $G_{11}=l G_{10}$). In other words, the reduced solution with D0 branes inherits features of the KKW/M2/M5 solution \rf{WM2M5int3}. On the other hand, the thermodynamics of the latter coincides with thermodynamics of the boosted M5's solution of Sec. 2.1.1. Therefore, one
can tune the parameters so as to get
the $n^3$ entropy growth in the near-extremal limit.

\subsection{Black branes solutions with the $n^3$ entropy growth in $9 \ge D \ge 6$}

\setcounter{equation}{0}

By dimensionally reducing the solutions of $n^3$ entropy, one may construct other near-extremal black branes with the $n^3$ scaling in lower dimensions. Here we focus on the solution \rf{D4E}, and obtain 9D D3, 8D D2, 7D D1 and 6D D0 black branes from it. The low-dimensional descents of D0/D2/D4 interpolating solution \rf{D0D2D4int}--\rf{D0D2D4phi} can be obtained in a similar way.

To construct descents of \rf{D4E} by toroidal dimensional reduction to 9D, 8D, 7D and 6D, we use the standard $D$-dimensional Kaluza-Klein ansatz
\be
g^{(D)}_{MN} dx^M dx^N=e^{2\a \phi_{(D-1)}} g^{(D-1)}_{mn}dx^m dx^n+e^{2\b \phi_{(D-1)}}\left(dx^\flat+A_m^{(D-1)}dx^m \right)^2 .
\la{KKans D10D7}
\ee
In the Einstein frame, $\a$ and $\b$ are given by
\be
\a=-\sqrt{\fr1{2(D-2)(D-3)}},\qquad \b=-(D-3)\a,\qquad D=10,9,8,7
\la{abD10D7}
\ee
Explicitly, they are given by
\[
\a=-\fr1{4\sqrt{7}},\qquad \b=\fr{\sqrt{7}}4, \qquad D=10 ;
\]
\[
\a=-\fr1{2\sqrt{21}},\qquad \b=\sqrt{\fr37}, \qquad D=9 ;
\]
\[
\a=-\fr1{2\sqrt{15}},\qquad \b=\fr12 \sqrt{\fr53}, \qquad D=8 ;
\]
\[
\a=-\fr1{2\sqrt{10}},\qquad \b=\sqrt{\fr25},\qquad D=7 .
\]
Proceeding in the standard manner and reducing along $x^\flat=x_4$, $x_3$, $x_2$, $x_1$
successively, one gets, for the graviton-dilaton part of solutions,
\begin{itemize}

\item
9D $n$D3 coincident black branes
\be
ds^2_9=H_3^{-3/7}\left(-fdt^2+dx_1^2+\dots+dx_3^2 \right)+H_3^{4/7}\left(f^{-1}dr^2+r^2 d\Omega^2_4 \right) ,
\la{D39D}
\ee
\be
e^{-\fr{4\sqrt{7}}{3}\phi_{(9)}}=H_3 \, ;
\la{D39Dphi}
\ee

\item
8D $n$D2 coincident black branes
\be
ds^2_8=H_2^{-1/2}\left(-fdt^2+dx_1^2+dx_2^2 \right)+H_2^{1/2}\left(f^{-1}dr^2+r^2 d\Omega^2_4 \right) ,
\la{D28D}
\ee
\be
e^{-2{\sqrt{\fr73}\phi_{(8)}}}=H_2 \, ;
\la{D28Dphi}
\ee

\item
7D $n$D1 coincident black branes
\be
ds^2_7=H_1^{-3/5}\left(-fdt^2+dx_1^2 \right)+H_1^{2/5}\left(f^{-1}dr^2+r^2 d\Omega^2_4 \right) ,
\la{D17D}
\ee
\be
e^{-2{\sqrt{\fr53}\phi_{(7)}}}=H_1 \, ;
\la{D17Dphi}
\ee

\item
6D $n$D0 non-extremal branes
\be
ds^2_6=-fH_0^{-3/4}dt^2 +H_0^{1/4}\left(f^{-1}dr^2+r^2 d\Omega^2_4 \right) ,
\la{D06D}
\ee
\be
e^{-2\fr{\sqrt{10}}{3} \phi_{(6)}}=H_0 \, .
\la{D06Dphi}
\ee
\end{itemize}

\ni Everywhere in \rf{D39D}--\rf{D06Dphi}
\be
H_i(r)=1+n\fr{h_0^3}{r^3},\qquad f(r)=1-\fr{\m^3}{r^3},\qquad i=3,2,1,0 \, ;
\la{Hf9876}
\ee
\be
F^{(D-1)}_{[4]}=3nh^3_0 \left(1+\fr{\m^3}{nh^3_0} \right)^{1/2} \e_4  , \qquad D-1=9,8,7,6\, .
\la{F9876}
\ee
Other scalars and tensors are set to zero.
For all of these systems, the ADM mass, entropy, and temperature are given by
\be
M_p=\fr{l^p \Omega_4}{16\pi G_{[p+6]}} \left[4\m^3+3n h_0^3 \right],
\la{Mp96}
\ee
\be
S_p=\fr{l^p \Omega_4}{4 G_{[p+6]}} \left(1+n\fr{h_0^3}{\m^3}\right)^{1/2} \m^4,
\la{Sp96}
\ee
\be
T_p=\fr3{4\pi}\left(1+n\fr{h_0^3}{\m^3}\right)^{-1/2} \m^{-1}.
\la{Tp96}
\ee
A comparison of these with \rf{D4M}--\rf{D4T} reveals that the solutions
\rf{D39D}--\rf{F9876}, in the near-extremal limit, have the $n^3$
near-extremal entropy scaling.



\section{Conclusions}

In this paper we have constructed a series of black brane solutions of 11D/10D IIA supergravities
 that exhibit the $n^3$ near-extremal entropy behavior with appropriately chosen scalings of the parameters. They include the boosted M5 black branes (M5/KKW bound state of \cite{Tseytlin:1996bh,Costa:1996re,Ohta:1997wp}) and the boosted M2/M5 interpolating black branes of \cite{Izquierdo:1995ms,Green:1996vh,Costa:1996re}. The identical near-extremal equation of state can be assigned to their dimensionally reduced versions, IIA D0/D4 black branes bound state and the solution of boosted D2/D4 interpolating black branes.

We have proposed a simple criterion for a solution to have the $n^3$ near-extremal entropy. The compelling candidates are those whose harmonic functions depend on five coordinates transverse to the host brane
woldvolume. In the spherical coordinate system it means that the harmonic functions will have $r^{-3}$-dependence.

Implementing dimensional reduction on the type IIA black D4 branes solution and taking the criterion as a guideline, we have established that black-brane solutions corresponding to 9D D3, 8D D2, 7D D1 and 6D D0 satisfy the criterion and all share the $n^3$ entropy growth. In fact, some of these solutions were previously known. For instance, the solution of 8D D2 black branes is similar to the purely magnetic D2 brane solution of \cite{Izquierdo:1995ms}.

Although some of the black-brane solutions constructed here share the $n^3$ entropy with the M5 solution of Klebanov-Tseytlin, the former differ from the latter in the following aspect. As noted in \cite{Klebanov:1996un}, the equations of state of non-dilatonic black branes in the near-extremal limit can be reproduced\footnote{This is with an exception of the M2 case for which one cannot find an agreement between the Bekenstein-Hawking and statistical entropy.} in terms of a weakly-interacting ideal gas of massless particles that are associated with massless excitation of $p$-brane modes. This is not the case for the descents of black M5 brane solutions constructed here, starting with the D4 black branes of 10D IIA. These descents are
{\em dilatonic} branes; non-perturbative corrections become important in $r \ra 0$ limit.

One novelty of the KKW/M2/M5 solution is thermodynamics ``duality''; there are two ways to scale the parameters such that
the thermodynamics of the two scalings toggles under the interchange of $n_5 \leftrightarrow n_0$. The novelty may be related to the fact that
it was constructed out of a M2/M5 solution that has only a {\em single} harmonic function. Perhaps the ``duality'' is an indication toward the supergravity analogue of the fact
that the SYM account of D-branes physics admit ``dual'' description, one in terms of
the lower dimensional branes and the other in terms of higher (see, e.g., \cite{Hatefi:2012sy} and refs therein).

We have also noticed that the ratio of the shear viscosity to entropy density of black M5 branes saturates the bound of the universal $\eta/S$ relation conjectured in \cite{Kovtun:2004de}. The same bound value is expected for the strongly-coupled 5D supersymmetric plasma at nonzero temperature due to the similarity of $S(T)$ near-extremal dependence and the hydrodynamical limit \cite{Natsuume:2008iy} shear viscosity of D4 branes with those of M5 branes. It would be interesting to establish the $\eta/S$ ratio for descents of D4 branes with the $n^3$ growth, and to verify the universal relation conjecture.

String corrections to the Einstein gravity action may provide additional information on the subleading terms in the universal $\eta/S$ relation of D4 branes, reflecting, in particular, the contribution of the hidden degrees of freedom of 5D supersymmetric gauge theory as well as its 6D cousin. Since a D4-brane is an example of non-conformal theory, string corrections on the SYM side will provide a new test on AdS/CFT correspondence. In our opinion, this is a test of fundamental importance, and deserves further studies. We hope to report on this and other issues in the future.

\ssk
\ssk
\ssk
\ni {\bf Acknowledgements}

\ni EH would like to thank  F. Quevedo, L. Alvarez-Gaume,
N. Lambert. He is grateful to A. Ghodsi for his very valuable insight and helpful conversations.
The work of AJN is supported in part by the Joint DFFD-RFBR Grant \#F40.2/040.
He thanks Vladimir Ivashchuk for the discussion on supergravity solutions. IP thanks Oleg Lunin for useful discussions.

\newpage
\renewcommand{\theequation}{A.\arabic{equation}}
 \setcounter{equation}{0}
  \section*{Appendix A: The ADM mass, the entropy and the Hawking temperature of black branes
  }

A most general line element used throughout the paper has the following form
\[
ds^2=- B^2(r)dt^2+C^2(r)\sum_{k=1}^{p-{\bar q}} dx^2_k+D^2(r)\sum_{{\bar i}=1}^{\bar q} dx^2_{\bar i}+E^2(r)\sum_{j=1}^{q-{\bar q}} dx^2_j
\]
\be
+F^2(r)dr^2+G^2(r) r^2 d\Omega^2_{d+1}  \, ,
\la{qpds}
\ee
 where $p+(q-\bar{q})+d+3=D$, $D$ is the number of space-time dimensions, $p$, $q$ and $\bar{q}$ are the number of spatial dimensions of intersecting $p$- and $q$-branes (we implicitly assume that $p\ge q$), and $\bar{q} \in \mathbb{Z}$ is the number of dimensions within the intersection. Configurations of coincident non-intersecting black branes correspond to $q=\bar{q}=0$. Functions $B(r),C(r),D(r),E(r),F(r),G(r)$ are some powers of the harmonic function(s) $H(r)=1+h^d/r^d$ of the space transverse to the $p$-brane \footnote{We choose the spherical coordinate system to parameterize the transverse space; the coordinates consist of the radial coordinate $r$ and the $d+1$ angles.}. $B(r)$, $F(r)$ also contain the ``black factor'' $f(r)=1-\mu^d/r^d$, setting the horizon at $r_H=\m$.

We are mainly interested in computing the entropy-temperature dependence in the near-extremal limit $\m \ll 1$. The entropy and the Hawking temperature for \rf{qpds}
 are given by (see, e.g., \cite{ArgurioPhD})\footnote{Recall that the Hawking temperature expression \rf{qpT} is correct only for metrics with flat space-time limit (under $r \ra \infty$). A case of a non-flat space-time limit and computation of its Hawking temperature can be found in \cite{Wu:2007sw}.}
\be
S=\fr{l^{[D-(d+3)]}\Omega_{d+1}}{4 G_D}C^{p-{\bar q}}D^{\bar q} E^{q-\bar{q}}(Gr)^{d+1} \vert_{r=r_H},
\la{qpS}
\ee
\be
T=\fr1{2\pi} \sqrt{g^{rr}}\fr{d}{dr} \sqrt{-g_{tt}}~ \vert_{r=r_H} .
\la{qpT}
\ee
In addition, one needs to compute the ADM mass (see, e.g., \cite{Horowitz:1991cd,Lu:1993vt})
\[
M=\fr{l^{[D-(d+3)]}\Omega_{d+1}}{16\pi G_D} \left[ (d+1)r^d (F^2-G^2)-(d+1)r^{d+1}(G^2)^\prime \right.
\]
\be
\left. -r^{d+1}\{ (p-{\bar q})(C^2)^\prime+\bar{q}(D^2)^\prime+(q-{\bar q})(E^2)^\prime \} \right] _{\vert r \ra \infty} .
\la{qpADM}
\ee

\renewcommand{\theequation}{B.\arabic{equation}}
\setcounter{equation}{0}

\section*{Appendix B: Basics of toroidal dimensional reduction
}

We give a brief summary of toroidal dimensional reduction from $D$ to $(D-1)$-dimensional space-time (see, e.g., \cite{Bergshoeff:1995as,Lu:1995cs,Bandos:2003et}).

The Kaluza-Klein ansatz for a $D$-dimensional metric is
\be
ds^2_D\equiv g^{(D)}_{MN} \, dx^M  dx^N=e^{2\a \phi_{(D-1)}} g_{mn}^{(D-1)}dx^m  dx^n+e^{2\b\phi_{(D-1)}}\left(dx^\flat+A^{(D-1)}_m dx^m\right)^2 ,
\la{KKansD}
\ee
 where $M,N,\dots$ run over $0,1,\dots,D-1$; $m, n,\dots$ run over $0,1,\dots, D-2$; $x^\flat$ is the direction of the reduction. $\phi_{(D-1)}$ and $A^{(D-1)}_m$ are the dilaton and the Kaluza-Klein vector field in $(D-1)$ space-time dimensions. They come from $D$-dimensional metric $g^{(D)}_{MN}$. The signature of the KK metric \rf{KKansD} is chosen to be mostly positive.

Dimensional reduction of the Einstein part of a $D$-dimensional supergravity action in the Einstein frame leads to
\[
\fr1{2k^2_D}\int \, d^D x \sqrt{-g^{(D)}}R^{(D)}=\fr1{2k^2_{D-1}} \int \, d^{D-1}x e^{((D-3)\a+\b)\phi}\sqrt{-g^{(D-1)}} \times
\]
\be
\times \left[ R^{(D-1)}+\a(D-2)\left[\a(D-3)+2\b \right] (\pa \phi)^2-\fr14 e^{2(\b-\a)}\left( F^{(D-1)}_{[2]}\right)^2 \right] ,
\la{RDEins}
\ee
\be
F^{(D-1)}_{[2]}=dA^{(D-1)}_{[1]}.
\la{KKF}
\ee
Both to get the gravity part of $(D-1)$-dimensional action and to reproduce the right factor $-1/2$ in front of the $(D-1)$-dimensional dilaton kinetic term, one fixes
\be
(D-3)\a+\b=0, \qquad \a^2=\fr1{2(D-2)(D-3)} \, .
\la{abE}
\ee
In the string frame, $(D-1)$ reduced action is associated with the following choice
\be
(D-3)\a+\b=-2,\qquad \a(D-2)\left[4+\a(D-3)\right]=-4.
\la{abS}
\ee
To avoid ambiguities we always choose the negative $\a$ first, and fix the corresponding $\b$ afterwards.

On account of the KK ansatz \rf{KKansD}, it is easy to recover the dimensionally reduced M-brane solutions. Let us describe the first descent of $D=11$ $n$ M5 black brane solution as an example; other lower dimensional brane configurations considered in the paper can be obtained in a similar manner.

Start with the M5s non-extremal solution \rf{M5Bng}
\be
ds^2_{11}=H^{-1/3}(r)\left[-f(r)dt^2+dx_1^2+\dots+dx_5^2 \right]+H^{2/3}(r)\left(\fr{dr^2}{f(r)}+r^2 d\Omega^2_4 \right),
\la{M5B}
\ee
and substitute it into the l.h.s. of \rf{KKansD}. Fixing the reduced coordinate to be $x_5=x^\flat$ (which corresponds to the double dimensional reduction of M5 branes), one gets
\be
e^{2\b \phi_{(10)}}=H^{-1/3}, \qquad A^{(10)}_{[1]}=0 .
\la{phiD-1}
\ee
Another part of the ten-dimensional solution can be recovered from
\be
ds^2_{10}=e^{-2\a\phi_{(10)}} \left[H^{-1/3}(r)\left[-f(r)dt^2+dx_1^2+\dots+dx_4^2 \right]+H^{2/3}(r)\left(\fr{dr^2}{f(r)}+r^2 d\Omega^2_4 \right) \right] .
\la{M5red}
\ee
From \rf{abE}, \rf{abS}, \rf{phiD-1}, one gets
\be
ds^2_{10}=H^{-3/8}\left[-f(r)dt^2+dx_1^2+\dots+dx_4^2 \right]+H^{5/8}(r)\left(\fr{dr^2}{f(r)}+r^2 d\Omega^2_4 \right),
\la{D4Eins}
\ee
\be
e^{\phi_{(10)}}=H^{-1/4}
\la{D4Einsphi}
\ee
in the Einstein frame, and
\be
ds^2_{10}=H^{-1/2}\left[-f(r)dt^2+dx_1^2+\dots+dx_4^2 \right]+H^{1/2}(r)\left(\fr{dr^2}{f(r)}+r^2 d\Omega^2_4 \right),
\la{D4Str}
\ee
\be
e^{\phi_{(10)}}=H^{-1/4}
\la{D4Strphi}
\ee
in the string frame. Harmonic functions $H(r)$ and $f(r)$ are the same as in \rf{M5BnH}.
The reduction of the gauge sector of the M5s solution \rf{F4M5Bn} goes as follows
\be
C^{(11)}_{[3]}=C^{(10)}_{[3]}+B^{(10)}_{[2]}\wg dx^\flat \ra F^{(11)}_{[4]}=F^{(10)}_{[4]}+F^{(10)}_{[3]}\wg \left( dx^\flat +A_{[1]}^{(10)} \right) ,
\ee
\[
F^{(10)}_{[4]}=dC^{(10)}_{[3]}-F^{(10)}_{[3]}\wg A_{[1]}^{(10)} ,\qquad F^{(10)}_{[3]}=dB^{(10)}_{[2]} ,
\]
and leads, with non-trivial ansatz for $C^{(10)}_{[3]}$ gauge field, to
\be
F^{(10)}_{[4]}=3h^3_0 \left(1+\fr{\m^3}{h^3_0} \right)^{1/2} \e_4.
\la{D4F4App}
\ee
Solutions \rf{D4Eins}--\rf{D4Strphi}, \rf{D4F4App} determine the background of 10D $n$ coincident black D4 ranes in different supergravity frames (cf., e.g., \cite{ArgurioPhD}).

\end{document}